\documentclass{IEEEtran}

\usepackage{graphicx}
\usepackage{color}
\usepackage{cite}
\usepackage{bbm}
\usepackage{amsmath}
\usepackage{dcolumn}
\usepackage{bm}
\usepackage{amsmath}
\usepackage{esint}
\PassOptionsToPackage{normalem}{ulem}
\usepackage{ulem}
\usepackage{amssymb}
\usepackage{overpic}
\usepackage{epic}
\usepackage{epsfig}
\usepackage{epstopdf}
\usepackage{soul}

\def\e{\begin{equation}}
\def\f{\end{equation}}
\def\_#1{{\bf #1}}

\def\.{\cdot}

\def\Re{{\rm Re\mit}}

\def\l#1{\label{#1}}
\def\r#1{(\ref{#1})}
\def\=#1{\overline{\overline{#1}}}

\def\aeeb{\={\alpha}_{\rm ee}}
\def\aemb{\={\alpha}_{\rm em}}
\def\ameb{\={\alpha}_{\rm me}}
\def\ammb{\={\alpha}_{\rm mm}}
\def\aeeo{\alpha_{\rm ee}^{\rm co}}
\def\aeer{\alpha_{\rm ee}^{\rm cr}}
\def\aemo{\alpha_{\rm em}^{\rm co}}
\def\aemr{\alpha_{\rm em}^{\rm cr}}
\def\ameo{\alpha_{\rm me}^{\rm co}}
\def\amer{\alpha_{\rm me}^{\rm cr}}
\def\ammo{\alpha_{\rm mm}^{\rm co}}
\def\ammr{\alpha_{\rm mm}^{\rm cr}}

\def\n{\eta_0}
\def\=#1{\overline{\overline{#1}}}

\def\It{\={I}_{\rm t}}
\def\Jt{\={J}_{\rm t}}

\begin{document}

\title{Least Upper Bounds of the Powers Extracted and Scattered by Bi-anisotropic Particles}

\author{I{\~n}igo~Liberal,
Younes~Ra'di,~\IEEEmembership{Student Member,~IEEE,}
Ram{\'o}n~Gonzalo,~\IEEEmembership{Member,~IEEE,}
I{\~n}igo~Ederra,
Sergei~A.~Tretyakov~\IEEEmembership{Fellow,~IEEE}
and
Richard~W.~Ziolkowski,~\IEEEmembership{Fellow,~IEEE}
\thanks{I. Liberal, I. Ederra and R. Gonzalo are with the Electrical and Electronic Engineering Department,
Universidad P\'{u}blica de Navarra, Pamplona, Spain,
e-mail: inigo.liberal@unavarra.es}
\thanks{Y. Ra'di and S. A. Tretyakov are with the
Department of Radio Science and Engineering,
Aalto University, P.O. 13000, FI-00076, Aalto, Finland
e-mail: younes.radi@aalto.fi}
\thanks{R. W. Ziolkowski is with the Department of Electrical and Computer Engineering,
University of Arizona, Tucson, AZ, 85721 USA,
e-mail:ziolkowski@ece.arizona.edu}
\thanks{This work was supported in part by the Spanish Ministry of Science and Innovation, Direcci\'on General de Investigaci\'on y Gesti\'on del Plan Nacional de I+D+I, Subdirecci\'on General de Proyectos de Investigaci\'on, Project Nos. TEC2009-11995 and CSD2008-00066 and by NSF contract number ECCS-1126572.}
\thanks{Manuscript received \textcolor{red}{Feb. XX, 2014}; revised ...}}

\maketitle

\begin{abstract}
The least upper bounds of the powers extracted and scattered by bi-anisotropic particles are investigated analytically. A rigorous derivation for particles having invertible polarizability tensors is presented, and the particles with singular polarizability tensors  that have been reported in the literature are treated explicitly. The analysis concludes that previous upper bounds presented for isotropic particles can be extrapolated to bi-anisotropic particles. In particular, it is shown that neither non-reciprocal nor magnetoelectric coupling phenomena can further increase those upper bounds on the extracted and scattered powers. The outcomes are illustrated further with approximate circuit model examples of two dipole antennas connected via a generic lossless network.
\end{abstract}


\section{Introduction\label{sec:Introduction}}

Maximizing the interaction between electromagnetic fields and electrically small particles is a topic of fundamental interest in many branches of physics \cite{Book_Bohren} and engineering \cite{Book_Tretyakov2003,Book_BalanisAEE}.
For example, maximizing the absorbed power is a main goal of any receiving antenna.
In this regard, it has been recognized that a finite size particle can absorb much more power than the amount that is projected onto its geometrical cross-section \cite{Bohren1982}. Moreover, it has been shown that the effective area of a finite size receiving antenna is unbounded \cite{Harrington1958,Book_Harrington}.
However, superdirective antennas with subwavelength sizes are usually accompanied with prohibitively narrow bandwidths and/or ill-posed numerical solutions \cite{Book_VolakisSmall}.
In contrast, the least upper bound of absorbed power for a finite number of interacting multipoles has been derived \cite{Liberal2013b,Pozar2009}. Interestingly, this maximal value is equal to the incident power density projected onto the effective area of Harrington's maximal antenna gain \cite{Harrington1958,Book_Harrington}.
Since this is the bound that would be obtained by invoking reciprocity,
it can be concluded that the presence of non-reciprocal materials is not required
to optimize the absorbed power \cite{Liberal2013b}.

Naturally, there is also a correlation between the absorption and scattering processes that limits the amount of power absorbed by a device as a function of its visibility \cite{Liberal2013b,Andersen2005}. Specifically, while the ratio of the absorbed to scattered power can be arbitrarily large, it comes at the expense of decreasing the absorbed power \cite{Liberal2013b,Andersen2005}. When the absorbed power is maximized, the absorbed and scattered powers
are equal \cite{Liberal2013b,Liberal2013a}.

Contrarily, there are a large number of applications that require the maximum visibility of a given device, which is directly linked to the maximization of the extracted and scattered powers. These applications include radar-based monitoring systems (e.g., passive RFID tags \cite{Rao2005} and mechanical stress sensors \cite{Liberal2012e,Occhiuzzi2011}), as well as any application related to the enhancement of electromagnetic field - matter interactions.
Recent multipolar \cite{Liberal2013b} and circuit model \cite{Liberal2013a} studies suggest that the least upper bound of the extracted and scattered powers is four times the bound of the absorbed power. While this result is intuitively understood by using a circuit model representation in which the scattering resistance cannot be suppressed \cite{Liberal2013a}, the aforementioned analyses are restricted to structures in which each multipole interacts independently. Although this is, in fact, the case for spherical bodies and single-mode structures, the scope of these works is limited since any object with sharp edges will enable the presence of and coupling between the excited multipoles.

It is noted that the coupling between the elementary electric and magnetic dipolar excitations is possible even for electrically small particles. Far from being a theoretical curiosity, different structures with reciprocal and non-reciprocal magnetoelectric couplings have been proposed \cite{Book_Bianisotropic}.
Moreover, it has been recognized that ensembles of these bi-anisotropic particles
enable the design of ultra-thin electromagnetic absorbers \cite{Radi2013a} and polarization transformers \cite{Niemi2013}, as well as one-way transparent sheets \cite{Radi2013c}.
When acting as single (isolated) elements, these bi-anisotropic particles provide more flexibility in the design of zero backward, zero forward and zero total scattering designs \cite{Vehmas2013}. Inversely, one might consider whether the presence of the
associated magnetoelectric couplings might help to maximize the scattering from an electrically small particle. In particular, the power extracted by electrically small uniaxial bi-anisotropic particles under different illumination conditions has been studied in \cite{Radi2013b}. This analysis demonstrates that bianistropic particles provide more flexibility in tailoring the power extracted from a given incident field, and it also suggests that the corresponding balance of the powers involved might well increase its maximal value.

In order to clarify this and other related issues, we investigate in this paper the least upper bound of the extracted and scattered powers for arbitrary particles, providing the generalization of the bound presented in \cite{Liberal2013b,Liberal2013a} to particles in which there are couplings between the excited multipoles. To this end, the discussion is structured as follows: First, Section\,\ref{sec:PextPscat} introduces basic definitions of scattered, absorbed, and extracted powers. Then, electrically small bi-anisotropic particles with elementary electric and magnetic dipolar excitations are considered in Section\,\ref{sec:Dipolar}.
The same procedure is generalized next in Section\,\ref{sec:Multimode} to particles that excite an arbitrary (though finite) number of multipoles.
Several circuit models are then presented in Section\,\ref{sec:Circuit_Model} to further illustrate the results and provide additional physical insights.
To finalize the discussion, the main results are summarized in Section\,\ref{sec:Conclusions}.

\section{Extracted and Scattered Powers\label{sec:PextPscat}}

Consider a particle illuminated by an incident electromagnetic field, which, in turn, scatters a certain amount of the incident power coupled to it. The incident and scattered fields will be labeled, respectively, as
$(\mathbf{E}^{\rm i}, \mathbf{H}^{\rm i})$
and
$(\mathbf{E}^{\rm s}, \mathbf{H}^{\rm s})$
and the total field as
$(\mathbf{E}^{\rm t}, \mathbf{H}^{\rm t})
= (\mathbf{E}^{\rm i}+\mathbf{E}^{\rm s}, \mathbf{H}^{\rm s}+\mathbf{H}^{\rm s})$.
Then the power radiated by the scatterer, i.e., the scattered power, is \cite{Liberal2013b}:
\begin{equation}
P_{\rm scat} = \frac{1}{2}\oiint_{S_{\infty}}\mathrm{Re}\left\{
\mathbf{E}^{\rm s}\times\left(\mathbf{H}^{\rm s}\right)^{*}
\right\} \cdot\widehat{\mathbf{n}} \> dS
\label{eq:Pscat}
\end{equation}

\noindent
where $S_{\infty}$ is a closed surface at infinity. Let the closed surface $S$ completely enclose the particle. The absorbed power $P_{\rm abs}$ is defined as the amount of power dissipated within the particle, and it can be computed as the inward flux of the total Poynting vector field through $S$:
\begin{equation}
P_{\rm abs} = -\frac{1}{2}\oiint_{S}\mathrm{Re}\left\{ \mathbf{E}^{\rm t}\times\left(\mathbf{H}^{\rm t}\right)^{*}\right\} \cdot\widehat{\mathbf{n}} \> dS
\label{eq:Pabs}
\end{equation}

Recall that according to the optical theorem \cite{Newton1976,GeneralizedOT}, the extracted power is defined as the sum of the absorbed and scattered powers, i.e., $P_{\rm ext}=P_{\rm scat}+P_{\rm abs}$. It corresponds to the power depleted from the incident field \cite{Newton1976,GeneralizedOT}, and/or the rate at which the incident field does work on the charge distributions excited in the particle \cite{Liberal2013a}. By invoking Poynting's theorem, the extracted power $P_{\rm ext}$ is found as the negative of the flux of the cross-term Poynting vector field through $S$:
\begin{equation}
P_{\rm ext} = - \oiint_{S} {\mathbf S}_{\rm cross} \cdot\widehat{\mathbf{n}} \> dS
\label{eq:Pext}
\end{equation}

\noindent
where
\begin{equation}
{\mathbf S}_{\rm cross}=\frac {1}{2}{\rm Re}\left\{
\mathbf{E}^{\rm i}\times\left(\mathbf{H}^{\rm s}\right)^{*} + \mathbf{E}^{\rm s}\times\left(\mathbf{H}^{\rm i}\right)^{*}
\right\}
\label{eq:Scross}
\end{equation}
\vskip 0.1in

\section{Dipolar Particles \label{sec:Dipolar}}

Consider an electrically small particle whose response when illuminated by the incident field can be approximated by the electric and magnetic dipole moments $\left(\mathbf{p},\mathbf{m}\right)$. In such a case, the extracted (\ref{eq:Pext}) and scattered (\ref{eq:Pscat}) powers can be written as (see, e.g., \cite{Book_Novotny})
\begin{equation}
P_{\mathrm{ext}}=
\frac{\omega}{2}
\mathrm{Im}\left\{\left(\mathbf{p}\right)^{*}\cdot\mathbf{E}^{\rm i} + \left(\mathbf{m}\right)^*\cdot\mathbf{H}^{\rm i}\right\} \label{eq:Pext_polarizabilities}
\end{equation}
\begin{equation}
P_{\mathrm{scat}}=\frac{\omega}{2}\frac{k_{0}^{3}}{6\pi\varepsilon_{0}}\left(\left|\mathbf{p}\right|^{2}+\frac{\left|\mathbf{m}\right|^{2}}{\eta_{0}^{2}}\right)\label{eq:Pscat_polarizabilities}
\end{equation}

\noindent
where $\eta_0$ is the free-space impedance. If the response of the particle is approximately linear, the induced electric and magnetic dipoles are proportional to the incident field, and they can be computed via
a given polarizability tensor as:
\begin{equation}
\left[\begin{array}{c}
\mathbf{p}\\
\mathbf{m}/\eta_{0}
\end{array}\right]
=\overline{\overline{\alpha}}\cdot\left[\begin{array}{c}
\mathbf{E}^{i}\\
\eta_{0}\mathbf{H}^{i}
\end{array}\right]
\label{eq:PalfaE}
\end{equation}

\noindent where the polarizability matrix $\overline{\overline{\alpha}}$
\begin{equation}
\overline{\overline{\alpha}}
=\left[\begin{array}{cc}
\overline{\overline{\alpha}}_{\rm ee} & \overline{\overline{\alpha}}_{\rm em}/\eta_{0}\\
\overline{\overline{\alpha}}_{\rm me}/\eta_{0} & \overline{\overline{\alpha}}_{\rm mm}/\eta_{0}^{2}
\end{array}\right]
\l{p}\end{equation}

\noindent describes the
electric,
$\aeeb$, and
magnetic,
$\ammb$,
polarization processes, as well as the magnetoelectric coupling $(\aemb,\ameb)$.
In this manner, extracted and scattered powers can be written as
\vskip 0.02in
\begin{equation}
P_{\mathrm{ext}}=\left[\begin{array}{c}
\mathbf{E}^{\rm i}\\
\eta_{0}\mathbf{H}^{\rm i}
\end{array}\right]^{\dagger}\cdot\overline{\overline{Q}}_{\mathrm{ext}}
\cdot\left[\begin{array}{c}
\mathbf{E}^{\rm i}\\
\eta_{0}\mathbf{H}^{\rm i}
\end{array}\right]
\label{eq:Pext_Matrix}
\end{equation}
\vskip 0.05in
\begin{equation}
P_{\mathrm{scat}}=\left[\begin{array}{c}
\mathbf{E}^{\rm i}\\
\eta_{0}\mathbf{H}^{\rm i}
\end{array}\right]^{\dagger}\cdot\overline{\overline{Q}}_{\mathrm{scat}}
\cdot\left[\begin{array}{c}
\mathbf{E}^{\rm i}\\
\eta_{0}\mathbf{H}^{\rm i}
\end{array}\right]
\label{eq:Pscat_Matrix}
\end{equation}

\noindent where
\begin{equation}
\overline{\overline{Q}}_{\mathrm{ext}}=\frac{\omega}{2}\frac{1}{2}\left[\left(-j\overline{\overline{\alpha}}\right)+\left(-j\overline{\overline{\alpha}}\right)^{\dagger}\right]
\label{eq:Qext}
\end{equation}
\vskip 0.05in
\begin{equation}
\overline{\overline{Q}}_{\mathrm{scat}}=\frac{\omega}{2}\frac{k_{0}^{3}}{6\pi\varepsilon_{0}}\overline{\overline{\alpha}}^{\dagger}\cdot\overline{\overline{\alpha}}
\label{eq:Qscat}
\end{equation}
%
%
and the dagger symbol represents the Hermitian (conjugate transpose) operation.

In this matrix formulation, the extracted and scattered powers are computed via the matrices $\overline{\overline{Q}}_{\mathrm{ext}}$ and $\overline{\overline{Q}}_{\mathrm{scat}}$,
which describe the response of the scatterer to the incident field.
It is apparent from (\ref{eq:Qext})-(\ref{eq:Qscat}) that both $\overline{\overline{Q}}_{\mathrm{ext}}$ and $\overline{\overline{Q}}_{\mathrm{scat}}$ are Hermitian matrices, which ensures that $P_{\mathrm{ext}}$ and $P_{\mathrm{scat}}$ are real numbers.
Moreover, equation~(\ref{eq:Qscat}) reveals that $\overline{\overline{Q}}_{\mathrm{scat}}$ is a positive semidefinite matrix, so that $P_{\mathrm{scat}}$ is always greater than or equal to zero.
In contrast, $\overline{\overline{Q}}_{\mathrm{ext}}$ is not necessarily a positive semidefinite matrix as $P_{\mathrm{ext}}$ can be negative for active particles. However, $\overline{\overline{Q}}_{\mathrm{ext}}$ must be positive semidefinite for passive particles, which imposes some constraints on the polarizability matrix $\overline{\overline{\alpha}}$.
For example, when considering lossless particles, the extracted and scattered powers must be equal for any incident field. This constraint imposes the condition: $\overline{\overline{Q}}_{\mathrm{ext}}=\overline{\overline{Q}}_{\mathrm{scat}}$,
which, in turn, means that the polarizability matrix must satisfy the matrix equation
\begin{equation}
\frac{1}{2}\left[\left(-j\overline{\overline{\alpha}}\right)
+\left(-j\overline{\overline{\alpha}}\right)^{\dagger}\right]=
\frac{k_{0}^{3}}{6\pi\varepsilon_{0}}\overline{\overline{\alpha}}^{\dagger}
\cdot\overline{\overline{\alpha}}
\label{eq:Polarizability_Lossless}
\end{equation}

\subsection{Least Upper Bound of the Extracted and Scattered Powers\label{sec:General_Dipolar}}

Since the presence of losses damps the excitation of currents within the particle, it follows that the extracted and scattered powers are maximized when the particle is lossless. By introducing a number of algebraic manipulations, the least upper bound of the extracted and scattered powers from the condition (\ref{eq:Polarizability_Lossless}) imposed on the polarizability matrix of lossless particles can be derived.
To this end, let us define initially the auxiliary matrix
\begin{equation}
\overline{\overline{P}}=-j\frac{k_{0}^{3}}{6\pi\varepsilon_{0}}\overline{\overline{\alpha}}
\label{eq:P}
\end{equation}

\noindent so that the matrix equation (\ref{eq:Polarizability_Lossless}) can be more conveniently written as
\begin{equation}
\frac{1}{2}\left(\overline{\overline{P}}+\overline{\overline{P}}^{\dagger}\right)
=\overline{\overline{P}}^{\dagger}\cdot\overline{\overline{P}}
\label{eq:P_Lossless}
\end{equation}

Consider first the following lemma, whose proof can be found in Appendix A: \break
\vskip 0.01in
\noindent
\textit{Lemma II.1 \,} Let $\overline{\overline{Q}} \> \in \> \mathbb{C}^{N\times N}$ be a Hermitian matrix generated from a matrix $\overline{\overline{P}} \> \in \> \mathbb{C}^{N\times N}$ as follows
\begin{equation}
\overline{\overline{Q}}
=\frac{1}{2}\left(\overline{\overline{P}}+\overline{\overline{P}}^{\dagger}\right)
=\overline{\overline{P}}^{\dagger}\cdot\overline{\overline{P}}
\label{eq:Q_Lemma}
\end{equation}
Then, if $\overline{\overline{P}}$ is invertible, $\overline{\overline{Q}}$ can be unitarily diagonalized and all its eigenvalues have absolute value smaller than one. This means that  $\overline{\overline{Q}}$ can be written as follows
\begin{equation}
\overline{\overline{Q}}
=\overline{\overline{U}}\cdot\overline{\overline{D}}_{Q}\cdot\overline{\overline{U}}^{\dagger}
\label{eq:Q_Unitary_Diagonalized}
\end{equation}

\noindent where $\overline{\overline{U}}$ is a unitary matrix $\overline{\overline{U}}\cdot\overline{\overline{U}}^{\dagger}=\overline{\overline{I}}$, and $\overline{\overline{D}}_{Q}$ is a diagonal matrix, whose diagonal elements have the absolute value smaller than 1, i.e.,
\begin{equation}
\left|\left[\overline{\overline{D}}_{Q}\right]_{nn}\right|\leq 1 \> \> \> \forall n
\label{eq:D_Q_nn}
\end{equation}

\noindent
Next, since $\overline{\overline{U}}$ is a unitary matrix, one knows that its
columns form an orthonormal basis of $\mathbb{C}^{N}$, i.e., $\mathbf{u}_{n}^{\dagger}\cdot\mathbf{u}_{m}=\delta_{nm}$. Thus, any vector $\mathbf{a} \> \in \> \mathbb{C}^{N}$ can be written as
\begin{equation}
\mathbf{a}=\sum_{n}a_{n}\mathbf{u}_{n}
\end{equation}

\noindent
Therefore, the quantity $\mathbf{a}^{\dagger}\cdot\overline{\overline{Q}}\cdot\mathbf{a}$
can be written as
\begin{equation}
\mathbf{a}^{\dagger}\cdot\overline{\overline{Q}}^{}\cdot\mathbf{a}
=\sum_{n}\left[\overline{\overline{D}}_{Q}\right]_{nn}\left|a_{n}\right|^{2}
\end{equation}

\noindent which, in view of (\ref{eq:D_Q_nn}), leads to the least upper bound
\begin{equation}
\left|\mathbf{a}^{\dagger}\cdot\overline{\overline{Q}}\cdot\mathbf{a}\right|
\leq\sum_{n}\left|a_{n}\right|^{2}
=\mathbf{a}^{\dagger}\cdot\mathbf{a}
\label{eq:Q_Bound}
\end{equation}

\noindent
Finally, note that $\overline{\overline{Q}} = \frac{\omega}{2}\frac{6\pi\varepsilon_{0}}{k_{0}^{3}}\overline{\overline{Q}}_{\rm ext}$. Then introducing (\ref{eq:Q_Bound}) into (\ref{eq:Pext_Matrix}), one finds that the least upper bound of the power extracted (and, hence, scattered) by a bi-anisotropic particle that has an invertible polarizability tensor is:
\begin{equation}
P_{\mathrm{ext}}\leq
\frac{\omega}{2}\frac{6\pi\varepsilon_{0}}{k_{0}^{3}}
\left(\left|\mathbf{E}^{\rm i}\right|^{2}+\eta_{0}^{2}\left|\mathbf{H}^{\rm i}\right|^{2}\right)
\label{eq:Pext_Bound}
\end{equation}
\vskip 0.1in

\noindent
From a mathematical standpoint, the bound (\ref{eq:Pext_Bound}) is a function of the incident field; it holds in both far-field and near-field scenarios.  However, one must be careful about the significance of $P_{\rm ext}$ in any near-field scenario where the power supplied by the sources is a function of the scattered field \cite{Liberal2013b}.

It is also convenient to particularize equation\,(\ref{eq:Pext_Bound}) to those incident fields in which the electric and magnetic field intensities are related through the medium impedance, i.e., as
$\left|\mathbf{E}^{\rm i}\right|^{2}=\eta_{0}^{2}\left|\mathbf{H}^{\rm i}\right|^{2}$.
This occurs, for example, for an incident plane-wave. In this case one is led to the expression:
\begin{equation}
P_{\mathrm{ext}}\leq\omega\frac{6\pi\varepsilon_{0}}{k_{0}^{3}}
\left|\mathbf{E}^{\rm i}\right|^{2}
= 4 \> \frac{\left|\mathbf{E}^{\rm i}\right|^{2}}{2\eta_{0}}
\left[ \> \frac{3 \> \lambda^{2}}{4\pi} \> \right]
\label{eq:Pext_Bound_PW}
\end{equation}

\noindent
Thus, the least upper bound is four times the incident power density projected onto the maximal effective area of a Huygens source antenna, i.e., one in which the electric and magnetic dipoles are balanced. This is the bound that could have been derived by invoking reciprocity. Therefore, it can be concluded that neither non-reciprocal effects nor magnetoelectric couplings are needed to optimize the power extracted and/or scattered by electrically small particles.

\subsection{Singular Polarizability Matrices and Balanced Particles}\label{sec:Balanced}

\tabcolsep=0.13cm
\begin{table*}[t]
\caption{Basic classes of uniaxial bi-anisotropic particles, magnetoelectric couplings, and optimal incident fields
\label{tab:Bi_anisotropic_classes}}
\begin{centering}
\begin{tabular}{cccc}
Omega & Chiral & Moving & Tellegen
\\
\hline
\\
$\overline{\overline{\alpha}}_{\rm em}=\overline{\overline{\alpha}}_{\rm me}=j\Omega\overline{\overline{J}}_{\rm t}$ & $\overline{\overline{\alpha}}_{\rm em}=-\overline{\overline{\alpha}}_{\rm me}=j\kappa\overline{\overline{I}}_{\rm t}$ & $\overline{\overline{\alpha}}_{\rm em}=-\overline{\overline{\alpha}}_{\rm me}=V\overline{\overline{J}}_{\rm t}$ & $\overline{\overline{\alpha}}_{\rm em}=\overline{\overline{\alpha}}_{\rm me}=\chi\overline{\overline{I}}_{\rm t}$
\\[1.75ex]
$\begin{array}{c}
\mathbf{E}^{\rm i}=E_{0}\left(\mathbf{x}+\mathbf{y}\right)\mathrm{cos}\left(k_{0}\left(z+\triangle z\right)\right)
\\[0.75ex]
\mathbf{H}^{\rm i}=j\frac{E_{0}}{\eta_{0}}\left(\mathbf{x}-\mathbf{y}\right)\mathrm{sin}\left(k_{0}\left(z+\triangle z\right)\right)
\end{array}$
&
$\begin{array}{c}
\mathbf{E}^{\rm i}=E_{0}\left(\mathbf{x}\pm j\mathbf{y}\right)e^{-jk_{0}z}\\[0.75ex]
\mathbf{H}^{\rm i}=\frac{E_{0}}{\eta_{0}}\left(\mathbf{y}\mp j\mathbf{x}\right)e^{-jk_{0}z}
\end{array}$
&
$\begin{array}{c}
\mathbf{E}^{\rm i}=E_{0}\left(\mathbf{x}+\mathbf{y}\right)e^{-jk_{0}z}\\[0.75ex]
\mathbf{H}^{\rm i}=\frac{E_{0}}{\eta_{0}}\left(\mathbf{y}-\mathbf{x}\right)e^{-jk_{0}z}
\end{array}$
& $\begin{array}{c}
\mathbf{E}^{\rm i}=E_{0}\left(\mathbf{x}\pm j\mathbf{y}\right)\mathrm{cos}\left(k_{0}\left(z+\triangle z\right)\right)
\\[0.75ex]
\mathbf{H}^{\rm i}=j\frac{E_{0}}{\eta_{0}}\left(\mathbf{y}\mp j\mathbf{x}\right)\mathrm{sin}\left(k_{0}\left(z+\triangle z\right)\right)
\end{array}$
\\
\end{tabular}
\par\end{centering}
\end{table*}

The proof presented in the previous section is restricted to particles with invertible polarizability matrices. It would appear that we are finished, at first glance, since
any polarizability matrix should in fact be invertible, otherwise the existence and uniqueness theorems of the inverse scattering problem would be violated.
However, since the polarizability matrices considered thus far, as defined by the assumption of electrically small particles, only contain their dipolar responses, it can be understood that such a matrix, in fact, can be singular. In particular, this can occur provided that the extended matrix which includes all of the higher-order modes (these being negligible in comparison to the dipolar modes in the computation of the extracted power)
is non-singular.
Moreover, because a singular polarizability matrix has non-zero nullity, there are incident field configurations for which the particles produce zero-scattering. Furthermore, these arrangements can only be achieved with lossless particles.

While these singular polarizability matrix configurations are theoretical limiting cases, the balanced uniaxial bi-anisotropic particles presented in \cite{Radi2013b} constitute examples that might optimize the extracted power. These singular cases are addressed next  to prove that they do not constitute counterexamples to the non-singular-based upper-bound (\ref{eq:Pext_Bound}).

Following \cite{Radi2013b} we consider uniaxial particles taken with respect to the z-axis, i.e., cases in which there is only excitation of electric and magnetic dipole moments along $\mathbf{x}$ and $\mathbf{y}$ and not along $\mathbf{z}$. This allows us to write the polarizability tensors as:
\begin{equation}
\aeeb=\aeeo\It+\aeer\Jt\label{eq:alpha_ee_dyadic}
\end{equation}
\begin{equation}
\ammb=\ammo\It+\ammr\Jt\label{eq:alpha_mm_dyadic}
\end{equation}
\begin{equation}
\aemb=\aemo\It+\aemr\Jt\label{eq:alpha_em_dyadic}
\end{equation}
\begin{equation}
\ameb=\ameo\It+\amer\Jt\label{eq:alpha_me_dyadic}
\end{equation}

\noindent where $\It=\mathbf{x}\mathbf{x}+\mathbf{y}\mathbf{y}$ is the transverse unit dyadic,  $\Jt=\mathbf{y}\mathbf{x}-\mathbf{x}\mathbf{y}$ is the dyadic defined in terms of the vector product operator,
and $\It = \mathbf{z} \times \Jt$ .

Bi-anisotropic particles are usually classified with respect to reciprocity and the symmetry of the magnetoelectric couplings:  $\aemb,\ameb$.
We consider here the classic omega, chiral, moving and Tellegen bi-anisotropic particles \cite{Book_Bianisotropic}, whose magnetoelectric couplings are defined in the first row of Table\,\ref{tab:Bi_anisotropic_classes}.
As identified in \cite{Radi2013b}, each of these balanced bi-anisotropy classes interacts optimally with specific incident field excitations, as exemplified in the second and third rows of Table\,\ref{tab:Bi_anisotropic_classes}. Specifically, linearly and circularly polarized propagating plane-waves optimally excite moving and chiral particles, respectively. On the other hand, linearly and circularly polarized standing plane-waves optimally excite omega and Tellegen particles, respectively.

For any of these bi-anisotropy classes, a lossless \textit{balanced} bi-anisotropic particle is defined as a particle satisfying \cite{Radi2013b}
\begin{equation}
\n\aeeo=j^{(\, c^2+1 \,)}\alpha_{\rm em}=\frac{\ammo}{\n}=-jr\label{eq:Balanced_Definition}
\end{equation}

\noindent with $\alpha_{\rm em}$ being any of the magnetoelectric parameters ($\alpha_{\rm em}^{\rm co}$ or $\alpha_{\rm em}^{\rm cr}$). Here $c=0,\-1,\-2,\-3$ correspond to chiral, Tellegen, omega, and moving particles, respectively. Moreover, we can assume that particles are at a resonance so that  $r\in \mathbb{R}$. It can be readily checked that if $r$ is a complex value, the only consequence is a reduction of the power extracted by such
a non-resonant particle. It also can be checked that balanced particles with zero cross-polar electric and magnetic polarizability terms (i.e., $\aeer = \ammr = 0$) result in singular polarizability matrices.

Power conservation imposes restrictions on polarizabilities of lossless bi-anisotropic particles \cite{Belov2003}.  In fact, the power balance condition for lossless balanced particles uniquely defines the value of $r$ in (\ref{eq:Balanced_Definition}). However, it is not possible to find $r$ by solving (\ref{eq:Polarizability_Lossless}), because the matrices are not invertible. To find
$r$, one needs to equate the extracted and scattered powers assuming from the very beginning that the balanced relation for the polarizabilities \r{eq:Balanced_Definition} is satisfied.
Direct calculations of the extracted and scattered powers show that all four omega, chiral, moving and Tellegen balanced and lossless particles share the same $r$ value, equal to
\begin{equation}
r=\frac{3\pi\sqrt{\mu_0\epsilon_0}}{k_0^3}
\label{eq:r_explicit_value}
\end{equation}

\noindent
With this result, the power extracted by singular particles of each class can be found by introducing (\ref{eq:r_explicit_value}) into (\ref{eq:Balanced_Definition}), and then inserting the ensuing  (\ref{eq:Balanced_Definition}) into the first row of Table~\ref{tab:Bi_anisotropic_classes} to define the
explicit particle polarizabilities. Next, the resulting first row of Table~\ref{tab:Bi_anisotropic_classes} is introduced into (\ref{eq:Qext}) and used in conjunction with the corresponding second and third rows to evaluate $P_{\mathrm{ext}}$ from (\ref{eq:Pext_Matrix}).

Subsequently, it is found that for the balanced particles illuminated by linearly and circularly polarized propagating plane-wave fields (i.e., for moving and chiral particles, respectively), the extracted power is given by
\begin{equation}
P_{\mathrm{ext}}=4\left(\frac{\left|E_{0}\right|^{2}}{\eta_{0}}\right)\left[ \> \frac{3 \> \lambda^{2}}{4\pi} \> \right]\label{eq:Pext_Balanced_Moving_Chiral}
\end{equation}

\noindent
Bearing in mind that $\left|\mathbf{E}^i\right|^2=2\left|E_{0}\right|^{2}$
for this incident field, it is clear that the power extracted by
balanced moving and chiral particles illuminated by their optimal excitation fields is equal to the upper bound (\ref{eq:Pext_Bound_PW}). On the other hand, for the balanced particles illuminated by linearly and circularly polarized standing plane-wave fields (i.e., for omega and Tellegen particles, respectively), the extracted power is given by
\[
P_{\mathrm{ext}}=\frac{\omega}{2}\frac{6\pi\varepsilon_{0}}{k_{0}^{3}}\left|E_{0}\right|^{2}\left[1+2\mathrm{sin}\left(k_{0}\triangle z\right)\mathrm{cos}\left(k_{0}\triangle z\right)\right]
\]
\begin{equation}
\leq\frac{\omega}{2}\frac{6\pi\varepsilon_{0}}{k_{0}^{3}}\left(2\left|E_{0}\right|^{2}\right)
\end{equation}

\noindent
Consequently, the balanced omega and Tellegen particles reach the upper bound given by equation (\ref{eq:Pext_Bound}) only when $k_{0}\triangle z=\frac{\pi}{4}+n\pi$.
Therefore, it can be concluded that the balanced particles presented in \cite{Radi2013b} can be considered optimal in the sense that they maximize the extracted and scattered powers
by reaching the upper bound of those magnitudes.

It is interesting to note that the values of the electric and magnetic polarizabilities
of the bi-anisotropic particle given by (\ref{eq:r_explicit_value}) are two times smaller than for the two disconnected dipolar particles. Nevertheless, this behavior is compensated by the active role of the magnetoelectric coupling coefficient. Therefore, the total power extracted by lossless resonant particles remains the same. It is also intriguing to note that the singular particles which maximize the extracted power for a given incident field  also produce zero scattering for other incident fields \cite{Vehmas2013}. Finally, it is important to remark that the upper bounds: (\ref{eq:Pext_Bound}) and (\ref{eq:Pext_Bound_PW}), can be reached with isotropic particles, i.e., no magnetoelectric coupling is needed to maximize those quantities.

\section{Multi-mode Particles \label{sec:Multimode}}

The previous dipolar results can be generalized for particles excited by an arbitrary (though finite) number of multipoles. To this end, we note that by adopting a multipolar formulation based on spherical harmonics, the extracted, scattered and absorbed powers can be written as \cite{Liberal2013b}
\footnote{Note that there is a sign change here in the definition of the
$A_{\left\{ q\right\} }$ coefficients with respect to \cite{Liberal2013b}
for a mere typographic convenience.}
\begin{equation}
P_{\mathrm{ext}}=\sum_{\left\{ q\right\} }\mathrm{Re}\left[A_{\left\{ q\right\} }^{*}B_{\left\{ q\right\} }\right]
\label{eq:Pext_Multipolar}
\end{equation}
\begin{equation}
P_{\mathrm{scat}}=\sum_{\left\{ q\right\} }\left|B_{\left\{ q\right\} }\right|^{2}
\label{eq:Pscat_Multipolar}
\end{equation}

\noindent where $A_{\left\{ q\right\} }$ and $B_{\left\{ q\right\} }$ stand for the incident and scattered field coefficients, and the multi-index $\left\{ q\right\} =\left\{ n,m,l,TZ\right\}$ is defined so that the series run over all spherical harmonics
\begin{equation}
\sum_{\left\{ q\right\} }=\sum_{n=1}^{\infty}\sum_{m=1}^{n}\sum_{l=e,o}\sum_{TZ=TM,TE}
\end{equation}

\noindent
It is worth noting that the same formulation can be applied to 2D structures \cite{Liberal2013c}. In such a case, the series would simply run over all cylindrical harmonics
\begin{equation}
\sum_{\left\{q\right\}}=\sum_{n=-\infty}^{\infty}\sum_{TZ=TM,TE}
\end{equation}

In order to construct a matrix formulation analogous to that introduced in Sec.\,\ref{sec:General_Dipolar}, let us gather all of our incident and scattered field multipolar coefficients into column vectors $\mathbf{A}$ and $\mathbf{B}$, respectively, with $\mathbf{A},\mathbf{B} \in \mathbb{C}^N$,
with $N$ being the number of multipoles.
In this manner,
Eqs.\,(\ref{eq:Pext_Multipolar})-(\ref{eq:Pscat_Multipolar})
can be rewritten as
\begin{equation}
P_{\mathrm{ext}}=\mathrm{Re}\left\{\mathbf{A}^{\dagger}\cdot\mathbf{B}\right\}
\label{eq:Pext_Matrices}
\end{equation}
\begin{equation}
P_{\mathrm{scat}}=\mathbf{B}^{\dagger}\cdot\mathbf{B}
\label{eq:Pscat_Matrices}
\end{equation}

\noindent
Again, when the response of the particle can be considered to be approximately linear, incident and scattering field coefficients can be related by a matrix:  $\overline{\overline{M}} \> \in \> \mathbb{C}^{N\times N}$, so that $\mathbf{B}=\overline{\overline{M}}\cdot\mathbf{A}$. Then the absorbed and scattered powers can be written in matrix form as follows
\begin{equation}
P_{\mathrm{ext}}
=\mathbf{A}^{\dagger}\cdot
\left[\frac{1}{2}\left(\overline{\overline{M}}
+\overline{\overline{M}}^{\dagger}\right)\right]\cdot\mathbf{A}
\label{eq:Pext_Sphe_Matrices}
\end{equation}
\begin{equation}
P_{\mathrm{scat}}
=\mathbf{A}^{\dagger}\cdot\left(\overline{\overline{M}}^{\dagger}
\cdot\overline{\overline{M}}\right)\cdot\mathbf{A}
\label{eq:Pscat_Sphe_Matrices}
\end{equation}

It is evident from (\ref{eq:Pext_Sphe_Matrices})-(\ref{eq:Pscat_Sphe_Matrices}) that the condition
$P_{\rm ext}=P_{\rm scat}$
for a lossless particle leads to the same matrix equation (\ref{eq:P_Lossless}). Therefore, the algebraic proof introduced in Sec.\,\ref{sec:General_Dipolar} can be directly applied to derive the following least upper bound for the power extracted by a multi-mode particle:
\begin{equation}
P_{\mathrm{ext}}\leq\mathbf{A}^{\dagger}\cdot\mathbf{A}
\label{eq:Pext_limit}
\end{equation}

\noindent
After notation changes, it can be readily checked that this least upper bound is equivalent to that derived in \cite{Liberal2013a,Liberal2013b} for particles in which each mode interacts independently, i.e., for a diagonal matrix $\overline{\overline{M}}$ in the present notation. Therefore, it is more generally concluded that no coupling between the distinct spherical modes is required to optimize the amount of power that can be extracted from the incident field.

\section{Circuit Models \label{sec:Circuit_Model}}

As a simple illustrative example, we consider the special case of two resonant dipole antennas (one electric and one magnetic) using the classical circuit model of receiving antennas. While these circuit models only provide an approximate solution to the scattering problem (see, e.g., \cite{Alu2010} and the references therein), they are a useful tool to illustrate and provide physical insight into the least upper bound derived in previous sections. Specifically, in order to illustrate the independence of the maximal extracted power from the manner in which the two antennas are connected, we compare the sum of the powers extracted by the two antennas when they are not connected, when they are directly connected to form a canonical chiral particle, and when they are connected with a generic network. With the use of the circuit model, we will be able to study the case of matched antennas, answer the question of maximizing the received power, and consider the case of the cloaked sensor regime (intentionally mismatched antennas with reduced scattered power).

\subsection{Disconnected and Directly Connected Antennas}

\subsubsection{Maximal extinction}

Let us consider a short electric dipole antenna (arm length $l$) and a small loop antenna (loop area $S$). The antennas are in the field of a circularly polarized propagating plane wave, and oriented so as to maximize their coupling to this
incident field. The incident fields are denoted: $E_{0}$ and $H_{0}=E_{0}/\eta_0$.
To find the maximum extinction cross section we make the antennas from PEC material and tune both antennas to resonance with bulk lossless reactive loads. The input impedances are then purely resistive and equal to the radiation resistances of the two antennas:
\e R_{\rm w}=\frac{\eta_0}{6\pi}(k_{0}l)^2,\qquad
R_{\rm l}=\frac{\eta_0}{6\pi}(k_{0}^2 S)^2\f

Let us assume that the polarizabilities are balanced so that the electromotive force induced in the wire antenna ${\cal E}=E_{0}l$ equals to that induced in the loop antenna ${\cal E}=\omega \mu_0 H_{0} S$. That is, we choose
$l=k_{0}S$. In this case the radiation resistances of the two antennas are also equal, and we denote them by $R$.

The powers extracted by these antennas from the incident fields are also equal. Consequently, the total power extracted by the two antennas is:
\begin{equation}
P_{\rm ext}=2\frac{1}{2}|I|^2 R =\frac{\left|{\cal E} \right|^2}{R}
\label{eq:max1}
\end{equation}

\noindent
Note that the currents in the antennas equal ${\cal E}/R$ because the antennas are loaded only by their radiation resistances. Substituting for the radiation resistance, we can check that the value of the maximal extinction cross section is at the upper bound given by (\ref{eq:Pext_Bound_PW}).

Let us next connect the two antennas together, forming one chiral particle. The electromotive force driving the current in the ``spiral'' is doubled, but also the load impedance is doubled. Thus, the extracted power is
\begin{equation}
P_{\rm ext}= \frac{4\left|{\cal E}\right|^2}{4R}R=\frac{\left|{\cal E} \right|^2}{R}
\label{eq:max2}
\end{equation}

\noindent which is the same as for the two individual (not connected) antennas (\ref{eq:max1}).

\subsubsection{Maximal received power}

This comparison can also be made for the case when we want to maximize the received power. This configuration is achieved by keeping the two antennas at resonance but loading them with matched loads. The received power is then the power delivered to these loads:
\e P_{\rm abs}=2{1\over 2} {\left|{\cal E} \right|^2\over 4R^2}R={\left|{\cal E} \right|^2\over 4R}\f

\noindent
The first factor of $2$ takes into account that we have two antennas. The total impedance of each is $2R$. We then calculate the power delivered to the resistor $R_{\rm load}=R$ in each dipole. If the two antennas are connected to form one chiral particle, we the obtain the same result:
\e P_{\rm abs}={1\over 2}{4\left|{\cal E} \right|^2\over (4R)^2} 2R={\left|{\cal E} \right|^2\over 4R}\f

\noindent
Now we have one antenna in which the induced electromotive force is doubled, but the radiation resistance is also doubled. To maximize the absorbed power, we load the antenna with the matched load $R_{\rm load}=2R$; and this reproduces the above result.

\subsubsection{Cloaked sensor regime}

Let us also compare the two cases in the regime where we want to increase the ratio of the absorbed and scattered powers, i.e., to make the scattering smaller and hide the sensor,
and determine how much power is received for a given ratio $P_{\rm abs}/P_{\rm scat}$.
To this end, we load each antenna with a resistor $R_{\rm load}$. Then the absorbed and scattered powers for the two antennas when they are disconnected are given by
\e P_{\rm abs}={\left|{\cal E} \right|^2\over (R+R_{\rm load})^2} R_{\rm load}\label{eq:Pabs_Cloaking_Disconnected}\f
\e P_{\rm scat}={\left|{\cal E} \right|^2\over (R+R_{\rm load})^2} R\label{eq:Pscat_Cloaking_Disconnected}\f

\noindent
Consequently, the absorbed to scattered power ratio is simply given by the impedance ratio
\e {P_{\rm abs}\over P_{\rm scat}}={R_{\rm load}\over R}\label{eq:PabsPscat_Cloaking_Disconnected}\f

\noindent
In this manner, in order to ensure a desired value of $P_{\rm abs}/P_{\rm scat}$, both antennas must be loaded with
\e R_{\rm load}=R{P_{\rm abs}\over P_{\rm scat}}\label{eq:Rload_Cloaking_Disconnected}\f

\noindent
Inversely, the absorbed power for a given value of the absorbed to scattered power ratio is given by
\e P_{\rm abs}={\left|{\cal E} \right|^2\over R}{{P_{\rm abs}\over P_{\rm scat}}\over \left(1+{P_{\rm abs}\over P_{\rm scat}}\right)^2}
\label{eq:Pabs_vs_Ratio_Cloaking_Disonnected}\f

Once again, if the two antennas are now simply connected together, there is a doubling
of the electromotive force and the resistance. Consequently, the absorbed and scattered powers are given by
\e P_{\rm abs}={1\over 2}{4\left|{\cal E} \right|^2\over (2R+R_{\rm load})^2} R_{\rm load}\label{eq:Pabs_Cloaking_Connected}\f
\e P_{\rm scat}={1\over 2}{4\left|{\cal E} \right|^2\over (2R+R_{\rm load})^2} 2R\label{eq:Pscat_Cloaking_Connected}\f

\noindent which leads to the absorbed to scattered power ratio expression:
\e {P_{\rm abs}\over P_{\rm scat}}={R_{\rm load}\over 2R}\label{eq:PabsPscat_Cloaking_Connected}\f

\noindent
It is then found that the load required to produce a certain absorbed to scattered power ratio in the connected case is twice the load required in the disconnected case (\ref{eq:Rload_Cloaking_Disconnected}):
\e R_{\rm load}=2R{P_{\rm abs}\over P_{\rm scat}}\label{eq:Rload_Cloaking_Connected}\f

\noindent
Despite this fact, it is found that the absorbed power obtained for a given absorbed to scattered power ratio is exactly the same as that obtained in the disconnected case (equation\,(\ref{eq:Pabs_vs_Ratio_Cloaking_Disonnected})):
\e P_{\rm abs}={\left|{\cal E} \right|^2\over R}{{P_{\rm abs}\over P_{\rm scat}}\over \left(1+{P_{\rm abs}\over P_{\rm scat}}\right)^2}
\label{eq:Pabs_vs_Ratio_Cloaking_Connected}\f

\noindent It now can be concluded that the same performance is obtained for the disconnected and simply connected antennas in the cloaked sensor regime.

\subsection{Non-reciprocal Connecting Networks}

The fundamental reason for the existence of the upper bound for the extracted power is the fact that any current element induced on any object by an external field creates secondary (scattered) fields. In terms of classic antenna circuit models, the antenna radiation resistance is the same both in reception and transmission. In terms of multipolar circuit models \cite{Liberal2013a}, the scattering resistance of each spherical mode is non-zero and independent of the properties of the scatterer. The inevitable presence of scattering losses damps the induced current in the antenna even if the antenna itself is lossless.

\begin{figure}
\centering
\epsfig{file=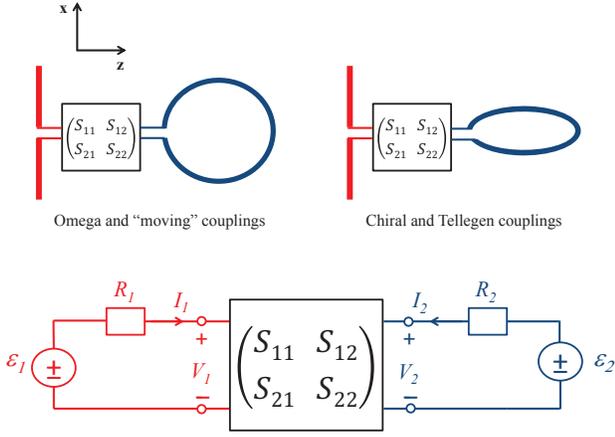, width=0.48\textwidth}
\caption{Equivalent circuit of a small dipole antenna and a loop antenna connected via a general two-port network.}
\label{fig:Network_Model}
\end{figure}

Here we analyze whether connecting two radiating elements with a general non-reciprocal network can break this rule. Consider a very general case of two small (dipolar) antennas connected through a two-port device characterized by a general scattering matrix (Fig.~\ref{fig:Network_Model}). The only assumption is that the two-port network is {\it lossless}, i.e., if we want to maximize the scattered power, the losses in the scatterer should be minimized. The circuits are the same models for the small wire and loop antennas used previously (assuming that both antennas have the same resonant frequency and both are at resonance, so that the input impedances are real). Conceptually, if the network could pass energy only from left to right, the antenna on the right would receive power from both sources but would scatter only through its own radiation resistance.

Solving the circuit model in Fig.\,\ref{fig:Network_Model} for a lossless network, it is found that the total power dissipated in the circuit (corresponding to the extracted power) is given by
\begin{equation}
\begin{array}{l}\displaystyle
P_{\rm ext}=P_1+P_2=\frac{1}{8}\left[\left(\frac{1}{R_1}+\frac{|S_{11}|^2}{R_1}+\frac{|S_{21}|^2}{R_2}\right)|{\cal E}_1|^2
\right.
\vspace{.2cm}\\\displaystyle
\left.
\hspace{.7cm}+\left(\frac{|S_{12}|^2}{R_1}+\frac{|S_{22}|^2}{R_2}+\frac{1}{R_2}\right)|{\cal E}_2|^2
\right.
\vspace{.2cm}\\\displaystyle
\left.
\hspace{.7cm}+2\Re \left\{\left(\frac{S_{11}S_{12}^*}{R_1}+\frac{S_{21}S_{22}^*}{R_2}\right){\cal E}_1{\cal E}_2^*
-\frac{S_{11}}{R_1}|{\cal E}_1|^2
\right.
\right.
\vspace{.2cm}\\\displaystyle
\left.
\left.
\hspace{.7cm}-\frac{S_{12}}{R_1}{\cal E}_1^*{\cal E}_2-\frac{S_{21}}{R_2}{\cal E}_1{\cal E}_2^*-\frac{S_{22}}{R_2}|{\cal E}_2|^2
\right\}
\right]\end{array}
\label{eq:Pext_Circuit}
\end{equation}

\noindent
Considering again that both the electric and magnetic dipoles are working in the balanced regime, i.e.,
when $|{\cal E}_1|=|{\cal E}_2|={\cal E}$ and $R_1=R_2=R$, the extracted power reads
\begin{equation}
P_{\rm ext}=\frac{|{\cal E}|^2}{2R}\left[1-\frac{1}{2}{\rm Re}\left\{S_{11}+j^{c}S_{12}+j^{-c}S_{21}+S_{22}\right\}\right]
\label{eq:Pext_Circuit_Balanced}
\end{equation}

\noindent
In this case, the term $c=0,\-1,\-2,\-3$ describes the different phase conditions of the products between the electromotive force induced at each port, ${\cal E}_{1}{\cal E}_{2}^*$, that correspond to the optimal excitations of the balanced chiral, Tellegen, omega and moving particles.

To prove that the extracted power in \r{eq:Pext_Circuit_Balanced} is equal to or less than ${|{\cal E}|^2}/{R}$, we must show that
\begin{equation}\begin{array}{l}\displaystyle
-\Re\left\{S_{11}+j^{c}S_{12}+j^{-c}S_{21}+S_{22}\right\}\le 2
\end{array}
\label{eq:Circuit_Condition_Bound}
\end{equation}

\noindent
To this end, we note that the scattering matrix of any lossless network is unitary:  $\overline{\overline{S}}\cdot\overline{\overline{S}}^\dagger = \overline{\overline{I}}$, which implies that
\begin{equation}
|S_{11}|=|S_{22}|\le 1, \quad |S_{12}|=|S_{21}|=\sqrt{1-|S_{11}|^2}
\label{eq:S_lossless1}
\end{equation}
It follows from (\ref{eq:S_lossless1}) that if no power can be dissipated inside the network, the absolute values of $S_{21}$ and $S_{12}$ are equal even in the most general non-reciprocal case.
Any matrix subject to (\ref{eq:S_lossless1}) can be written as
\begin{equation}
\overline{\overline{S}}=
\begin{array}{c}
\left[ \begin{array}{cc}
S_{11} &
S_{12}\\S_{21} &
S_{22} \end{array} \right]=
\left[ \begin{array}{cc}
e^{j\theta}\sin{\beta} &
e^{j\phi}\cos{\beta}\\e^{j\psi}\cos{\beta} &
e^{j\zeta}\sin{\beta} \end{array} \right]
\end{array}
\label{eq:S_lossless}
\end{equation}

\noindent where $\beta \in \mathbb{R}$. In addition to the above restrictions on the absolute values of the S-parameters, components of any unitary matrix satisfy
\e S_{11}S_{12}^*+S_{21}S_{22}^*=0\f
This imposes the following condition on the phases:
\begin{equation}
\theta+\zeta=\psi+\phi+(2n+1)\pi
\label{eq:hhh}
\end{equation}

\noindent
Introducing (\ref{eq:S_lossless}) into \r{eq:Circuit_Condition_Bound}, one obtains
\begin{equation}\begin{array}{l}\displaystyle
-\Re\left\{S_{11}+j^{c}S_{12}+j^{-c}S_{21}+S_{22}\right\}\vspace{0.2cm}\\\displaystyle
=-\sin{\beta}\left[\cos{\theta}+\cos{\zeta}\right]
\vspace{0.2cm}\\\displaystyle
\hspace{1cm}-\cos{\beta}\left[\cos\left(\phi+\frac{c\pi}{2}\right)+\cos\left(\psi-\frac{c\pi}{2}\right)\right]
\vspace{0.2cm}\\\displaystyle
\hspace{1cm}=2\sin{\beta}\cos\left(\frac{\theta+\zeta}{2}\right)\cos\left(\frac{\theta-\zeta}{2}+\pi\right)
\vspace{0.2cm}\\\displaystyle\hspace{1.3cm}
+2\cos{\beta}\cos\left(\frac{\phi+\psi}{2}\right)\cos\left(\frac{\phi-\psi}{2}+\frac{(c+2)\pi}{2}\right)
\end{array}
\label{eq:jjj}
\end{equation}

\noindent
The sums of the angles, $\theta+\zeta$ and $\phi+\psi$, in this expression are related by (\ref{eq:hhh}). On the other hand, the arguments of the last cosine functions in this sum can be arbitrary.
Clearly, the sum is maximized when
\begin{equation}\begin{array}{l}\displaystyle
\left|\cos\left(\frac{\theta-\zeta}{2}+\pi\right)\right|=\-\left|\cos\left(\frac{\phi-\psi}{2}+\frac{(c+2)\pi}{2}\right)\right|= 1
\end{array}
\label{eq:kkk1}
\end{equation}
It is important to note that if we make this assumption, we do not impose any conditions on the arguments of the other sine and cosine functions in (\ref{eq:jjj}).
Substituting (\ref{eq:kkk1}) into (\ref{eq:jjj}), we find
\begin{equation}\begin{array}{l}\displaystyle
-\Re\left\{S_{11}+j^{c}S_{12}+j^{-c}S_{21}+S_{22}\right\}\le\vspace{0.2cm}\\\displaystyle\hspace{0.5 cm}
2\sin{\beta}\cos\left(\frac{\theta+\zeta}{2}\right)
+2\cos{\beta}\cos\left(\frac{\phi+\psi}{2}\right)
\end{array}
\label{eq:jjj1}
\end{equation}
Next, using \r{eq:hhh} to express  $\theta+\zeta$ in terms of $\phi+\psi$, we re-write \r{eq:jjj1} as
\begin{equation}\begin{array}{l}\displaystyle
-\Re\left\{S_{11}+j^{c}S_{12}+j^{-c}S_{21}+S_{22}\right\}\vspace{0.2cm}\\\displaystyle
\hspace{.25cm}\le 2\left[-(-1)^h\sin{\beta}\sin\left(\frac{\phi+\psi}{2}\right)
+\cos{\beta}\cos\left(\frac{\phi+\psi}{2}\right)\right]
\vspace{0.2cm}\\\displaystyle
=2\cos\left(\beta+(-1)^h\frac{\phi+\psi}{2}\right)
\le 2
\end{array}
\label{eq:lll}
\end{equation}

\noindent
($h$ is an integer). This proves that the condition \r{eq:Circuit_Condition_Bound} is satisfied. Consequently,
no lossless connecting network can increase the bound: $\left|{\cal E}\right|^2/R$.

For example, note that the balanced bi-anisotropic particles of all four classes, whose polarizablities are given in (\ref{eq:Balanced_Definition}), can be realized as two antennas connected by a two-port with the matrix
\begin{equation}
\overline{\overline{S}}=
\begin{array}{c}
\left[ \begin{array}{cc}
S_{11} &
S_{12}\\S_{21} &
S_{22} \end{array} \right]=
\left[ \begin{array}{cc}
0 &
-(-j)^c\\-(-j)^{-c} &
0 \end{array} \right]
\end{array}
\label{eq:nnn}
\end{equation}
Substituting this $\overline{\overline{S}}$ matrix into \r{eq:Pext_Circuit_Balanced}, it is found that the extracted power obtains its maximum:
\begin{equation}
P_{\rm ext}=\frac{|{\cal E}|^2}{R}
\label{eq:ooo}
\end{equation}
This illustrative example clearly shows how it is not possible to suppress the scattering resistance of lossless connected particles even with a non-reciprocal connecting network. The reason resides in the restrictions associated with the scattering matrix of any lossless network \cite{Haus}.

\section{Conclusions\label{sec:Conclusions}}

This work has investigated the least upper bounds of the powers extracted and scattered by bi-anisotropic particles. Both electrically small (dipolar) particles and particles excited by an arbitrary number of spherical multipoles were considered. Since the presence of losses damps the currents excited in any structure, it is first concluded that both the extracted and scattered powers are maximized in an ideally lossless structure, and, therefore, that their least upper bound is the same.
Next, a rigorous derivation of the least upper bound for particles with invertible polarizability tensors was presented. While this result could be considered a general and complete proof, particles with singular polarizability tensors exist as theoretical limiting cases. Therefore, the previously reported singular cases of balanced omega, chiral, moving and Tellegen particles were treated individually. This collection of cases also represents a complete survey on the most popular magnetoelectric coupling effects.
The analysis concludes that in all of the aforementioned invertible and singular cases,
the least upper bound is equal to the bound previously reported for isotropic particles. Therefore, while the present analysis does not represent a complete and rigorous proof of the least upper bound of all bi-anisotropic particles (additional singular particles are likely to exist), we believe that the study gathers sufficient evidence to conjecture that, in contrast to a statement in \cite{Radi2013b}, neither non-reciprocal nor magnetoelectric coupling phenomena are required to maximize the extracted and scattered powers. Physically, the reason for this strict limit is that within this scenario, the scattering resistance of antenna elements cannot be suppressed.

From a practical standpoint, it is concluded that neither non-reciprocity nor bi-anisotropy are required to maximize the visibility of a given particle. This conclusion is of general interest for any radar-based sensing and/or imaging technology. However, it is also found that bi-anisotropic particles might provide additional functionalities while reaching the least upper bound of extracted power. For example, the studied singular particles provide the extreme behavior of maximizing the extracted power for a specific incident field, while producing zero-scattering for a different field excitation. This exotic behavior cannot be achieved with simple isotropic particles.

\appendices
\section{\label{sec:Proof}}

To prove \textit{Lemma II.1}, let us define the matrix
\begin{equation}
H\left(\overline{\overline{P}}\right)=\frac{1}{2}\left(\overline{\overline{P}}
+\overline{\overline{P}}^{\dagger}\right)
\label{eq:P_Hermitian}
\end{equation}

\noindent as the Hermitian part of $\overline{\overline{P}}=-j\frac{k_0^3}{6\pi\varepsilon_{0}}\overline{\overline{\alpha}}$. If the inverse of $\overline{\overline{P}}$ exists, we can also define the matrix
\begin{equation}
\overline{\overline{C}}=\left(\overline{\overline{P}}^{\dagger}\right)^{-1}\overline{\overline{P}}
\label{eq:Cayley_P}
\end{equation}

\noindent
as the generalized Cayley transform of $\overline{\overline{P}}^{\dagger}$ \cite{Fan1972}. A useful property of matrices with a positive definite Hermitian part is that their generalized Cayley transform is similar to a unitary matrix \cite{Fan1972,Mathias1992}. To exploit this fact, first note the property
\begin{equation}
\overline{\overline{C}}^{\dagger}\cdot H\left(\overline{\overline{P}}\right)\cdot\overline{\overline{C}}
=H\left(\overline{\overline{P}}\right)
\label{eq:CHC_H}
\end{equation}

\noindent
which can be checked by applying the definitions of $\overline{\overline{C}}$ and $H\left(\overline{\overline{P}}\right)$.
Next, introducing (\ref{eq:P_Lossless}) into (\ref{eq:CHC_H}) and multiplying on the left by $\left(\overline{\overline{P}}^{\dagger}\right)^{-1}$ and on the right by $\overline{\overline{P}}^{-1}$, it is found that
\begin{equation}
\begin{split}
\left(\overline{\overline{P}}^{\dagger}\right)^{-1}\cdot\overline{\overline{C}}^{\dagger}\cdot\overline{\overline{P}}^{\dagger}\cdot\overline{\overline{P}}\cdot\overline{\overline{C}}\cdot\overline{\overline{P}}^{-1}
\\
=\left(\overline{\overline{P}}\cdot\overline{\overline{C}}\cdot\overline{\overline{P}}^{-1}\right)^{\dagger}\cdot\left(\overline{\overline{P}}\cdot\overline{\overline{C}}\cdot\overline{\overline{P}}^{-1}\right)=\overline{\overline{I}}\label{eq:Unitary_Similar}
\end{split}
\end{equation}

\noindent
It can be concluded from (\ref{eq:Unitary_Similar})
that $\overline{\overline{C}}$ is similar to the unitary matrix $\overline{\overline{P}}\cdot\overline{\overline{C}}\cdot\overline{\overline{P}}^{-1}$. The latter can be written in more explicit form as
\begin{equation}
\overline{\overline{P}}\cdot\overline{\overline{C}}\cdot\overline{\overline{P}}^{-1}=\overline{\overline{P}}\cdot\left(\overline{\overline{P}}^{\dagger}\right)^{-1}
\end{equation}

\noindent
This relation reveals that $\overline{\overline{P}}\cdot\left(\overline{\overline{P}}^{\dagger}\right)^{-1}$
is indeed a unitary matrix. Consequently, it can be diagonalized as:
\begin{equation}
\overline{\overline{P}}\cdot\left(\overline{\overline{P}}^{\dagger}\right)^{-1}
=\overline{\overline{U}}\cdot\overline{\overline{D}}\cdot\overline{\overline{U}}^{\dagger}
\label{eq:Unitary_Diagonalized}
\end{equation}

\noindent
where $\overline{\overline{U}}$ is also a unitary matrix $\overline{\overline{U}}\cdot\overline{\overline{U}}^{\dagger}=\overline{\overline{I}}$, and $\overline{\overline{D}}$ is a diagonal matrix, whose diagonal elements have the absolute value 1, i.e.,
\begin{equation}
\left|\left[\overline{\overline{D}}\right]_{nn}\right|=1\, \>\> \forall n
\label{eq:Dnn}
\end{equation}

In order to exploit this property, let us multiply equation~(\ref{eq:P_Lossless}) by
$\left(\overline{\overline{P}}^{\dagger}\right)^{-1}$ and rearrange the terms to write
\begin{equation}
\overline{\overline{P}}^{\dagger}=\frac{1}{2}\left[\overline{\overline{I}}-\left(\overline{\overline{P}}\cdot\left(\overline{\overline{P}}^{\dagger}\right)^{-1}\right)^{-1}\right]
\label{eq:Q-1-1}
\end{equation}

\noindent where, in view of (\ref{eq:Unitary_Diagonalized}), can then be written as
\begin{equation}
\overline{\overline{P}}^{\dagger}
=\overline{\overline{U}}
\cdot\left[\frac{1}{2}\left(\overline{\overline{I}}
-\overline{\overline{D}}^{-1}\right)\right]
\cdot\overline{\overline{U}}^{\dagger}
\label{eq:Ptrans_Diagonalized}
\end{equation}

\noindent
Then, introducing (\ref{eq:Ptrans_Diagonalized}) and its conjugate transpose into (\ref{eq:Q_Lemma}), it is found that $\overline{\overline{Q}}$ can be diagonalized as follows
\begin{equation}
\overline{\overline{Q}}
=\overline{\overline{U}}\cdot\overline{\overline{D}}_{Q}\cdot\overline{\overline{U}}^{\dagger}
\label{eq:Q_Diagonalized_Appendix}
\end{equation}

\noindent with $\overline{\overline{D}}_{Q}$ being a diagonal matrix whose elements are given by
\begin{equation}
\left[D_{Q}\right]_{nn}=\frac{1}{2}\left(1-{\rm Re}\left\{ \left[D\right]_{nn}^{-1}\right\} \right)
\label{eq:D_Q_nn_Apendix}
\end{equation}

\noindent so that, in view of (\ref{eq:Dnn}), it follows
\begin{equation}
\left|\left[D_{Q}\right]_{nn}\right|\leq1\, \>\> \forall n
\label{eq:D_Q_nn_Apendix_Bound}
\end{equation}

\noindent
This completes the proof of Lemma II.1.

\bibliographystyle{IEEEtran}
\bibliography{IEEEabrv,library}

\end{document}